\documentclass[10pt,a4paper,twoside, twocolumn]{article}
\usepackage{graphicx}

\usepackage[T1]{fontenc}
\usepackage{amsmath}
\usepackage{caption}

\makeatletter
\renewcommand{\section}{\@startsection{section}{2}{0cm}{-\baselineskip}
{0,5\baselineskip}{\normalsize\bfseries}}
\renewcommand{\subsection}{\@startsection{subsection}{3}{0cm}{-\baselineskip}
{0,5\baselineskip}{\normalsize\slshape}}
\makeatother

\newcommand{\figwid}{7.6cm}
\newcommand{\doublefigwid}{7.9cm}
\newcommand{\plots}[1]{#1}

\renewcommand{\thefootnote}{\fnsymbol{footnote}} 

 \newcounter{ctr}

\begin{document}

\title{Qualification Tests of 474 Photomultiplier Tubes for the Inner Detector of the Double Chooz Experiment}
\author{C. Bauer$^a$, E. Borger$^a$, R. Hofacker$^a$, K. J\"{a}nner$^a$, F. Kaether$^{a*}$, C. Langbrandtner$^a$, \\ M. Lindner$^a$, S. Lucht$^{b}$\footnotemark[1], M. Reissfelder$^a$, S. Sch\"{o}nert$^a$,  A. St\"{u}ken$^{b}$, C. Wiebusch$^b$ \\}
\date{
 \small \sl 
 {$^a$}Max-Planck-Institut f\"{u}r Kernphysik, \\ 
 Saupfercheckweg 1, D-69117 Heidelberg Germany. \\
 E-mail: {\tt Florian.Kaether@mpi-hd.mpg.de} \\ 
 \vspace{0.2cm}
 {$^b$}RWTH Aachen, III. Physikalisches Institut B, \\
 Otto-Blumenthal-Stra\ss{}e, D-52056 Aachen, Germany \\
 E-mail: {\tt sebastian.lucht@physik.rwth-aachen.de} \\
 \vspace{0.4cm}
 {\small (Accepted by Journal of Instruments on 27 May 2011)}
}

\twocolumn[
\begin{@twocolumnfalse}
\maketitle
\begin{abstract}
\noindent The hemispherical 10'' photomultiplier tube (PMT) R7081 from Hamamatsu Photonics K.K. (HPK) is used in various experiments in particle and astroparticle physics. We describe the test and calibration of 474 PMTs for the reactor antineutrino experiment Double Chooz. The unique test setup at Max-Planck-Institut f\"{u}r Kernphysik Heidelberg (MPIK) allows one to calibrate 30 PMTs simultaneously and to characterize the single photoelectron response, transit time spread, linear behaviour and saturation effects, photon detection efficiency and high voltage calibration.\\
\end{abstract}
\end{@twocolumnfalse}
]

\footnotetext[1]{Corresponding authors}

\renewcommand{\thefootnote}{\arabic{footnote}}

\tableofcontents

\section{Introduction}

Double Chooz \cite{DCProposal} is a reactor antineutrino experiment located next to the nuclear power plant at Chooz, France. Its main physics goal is
to measure a non-vanishing value of the neutrino mixing angle $\theta_{13}$ or to improve the current upper limit given by a global analysis \cite{Schwetz}
\begin{equation}
\sin^2 (2\theta_{13}) \le 0.12 \, \; \mbox{at 90\% CL} \, .
\end{equation}
The two detector concept will substantially decrease systematic errors  and leads to a sensitivity of 
\begin{equation}
\sin^2 (2\theta_{13}) \le 0.03 \; \mbox{at 90\% CL}
\end{equation}

\begin{figure*}[t]
  \centering
  \includegraphics[width=0.34\textwidth]{\plots{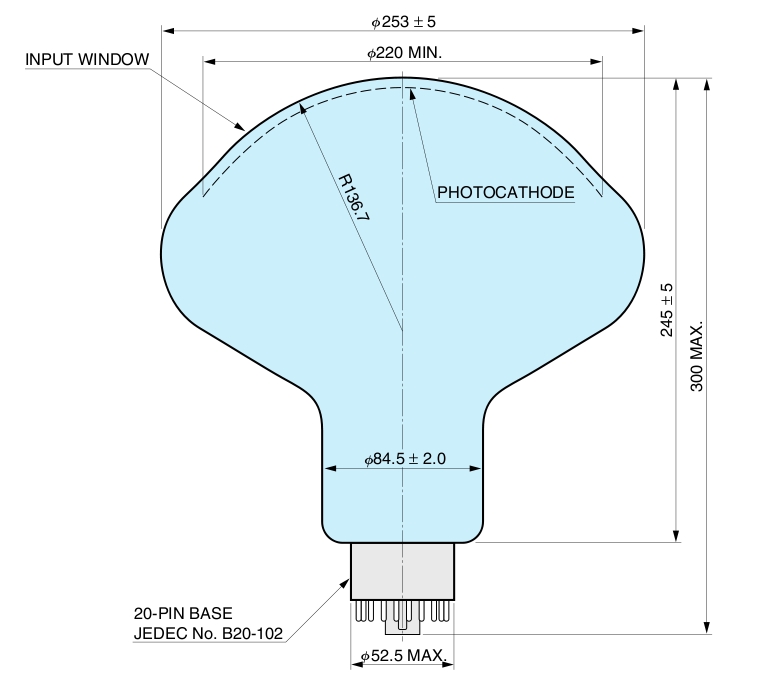}}
  \includegraphics[width=0.64\textwidth]{\plots{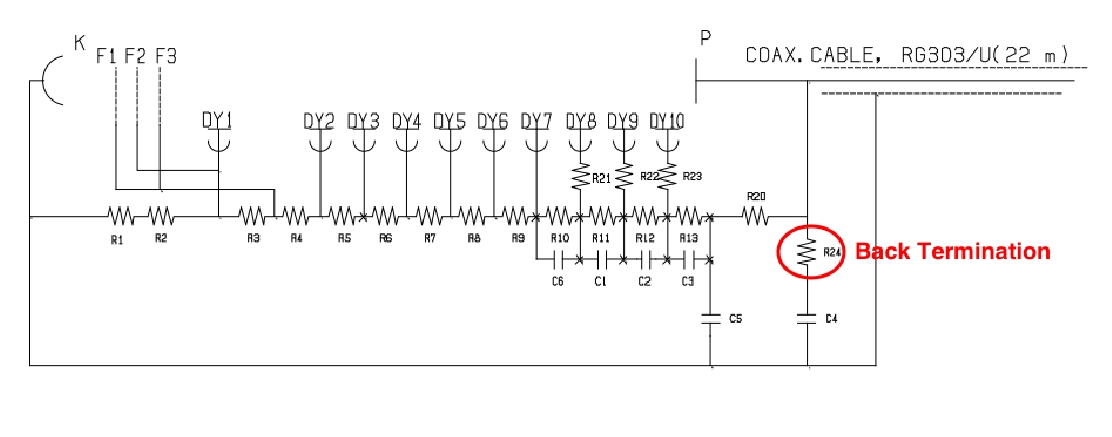}}
  \caption{The PMT Hamamatsu R7081 and the base circuit used for Double Chooz (taken from \cite{PMT} and \cite{japanese-pub}). R1, R2 = 2.5 M$\Omega$; R3 = 180 k$\Omega$; R4 = 1.02 M$\Omega$; R5 = 1.5 M$\Omega$; R6 = 1 M$\Omega$; R7 = 499 k$\Omega$; R8 = 300 k$\Omega$; R9 = 360 k$\Omega$; R10 = 430 k$\Omega$; R11 = 680 k$\Omega$; R12 = 910 k$\Omega$; R13 = 750 k$\Omega$; R20 = 10 k$\Omega$; R21-R23 = 100 $\Omega$; R24 = 49.9 $\Omega$; C1-C3, C6 = 10nF; C4, C5 = 4.7 nF.}
  \label{PMT-R7081}
\end{figure*}%

\begin{table}[t]
\begin{tabular}{|c|c|}
     \hline
     General parameters of R7081 & Description \\
     \hline
     Spectral Response  & 300 to 600 nm\\
     Peak Wavelength & 420 nm \\
     Material of photocathode & Bialkali \\
     Effective  photocathode area & 220 mm dia. \\
     Structure of dynodes & Box and Line\\
     Number of dynodes    & 10  \\
     Storage Temperature  & 0 to +40 ($^\circ$C)\\
     \hline
     \end{tabular}
     \caption{General parameters of the photomultiplier tubes Hamamatsu R7081 \cite{PMT}.}
     \label{PMTpara}
\end{table}

\noindent after 4 years of data taking \cite{DCProposal}. 
Each of the two inner detectors of Double Chooz will be observed by 390 photomultiplier tubes (PMTs) to record light pulses produced by neutrino-induced reactions inside the gadolinium-doped scintillator target volume.
The tasks of testing and calibrating these PMTs were shared between
German groups (MPIK Heidelberg and RWTH Aachen) and Japanese groups (see Acknowledgements). Both groups worked with independent calibration setups. However, careful
cross-checks between both setups showed all results in good agreement.
This article describes the results obtained by the German groups, and the test setup which was built at MPIK Heidelberg. The Japanese setup and results are described in \cite{japanese-pub}.

The PMTs for the inner Double Chooz detector are 10 inch in diameter and have a hemispherical surface with a bialkali photocathode (type R7081MOD-ASSY) from HPK \cite{PMT} (see Figure \ref{PMT-R7081} and Table \ref{PMTpara}). They are made of glass which is optimized for  radiopurity. 
The specifications of the PMTs (see Table \ref{PMTspecs}) have been validated for the purpose of a preselection for the Double Chooz experiment. Furthermore, detailed characterizations of the PMT behaviour are required for a precise understanding of the final detector response.

These aims lead to the following calibration procedure which is performed for every PMT:
\begin{list}{\arabic{ctr}.}{
  \usecounter{ctr}
  \setlength{\topsep}{1ex}
  \setlength{\itemsep}{-0.3ex}
  \setlength{\leftmargin}{3.5ex}
}
\item investigation of the dark count rate during and after the stabilization phase
\item determination of the high voltage (HV) value for a gain of $10^7$ per photoelectron
\item characterization of the single photon electron (SPE) response regarding spectral energy resolution and peak to valley ratio
\item exploration of relative transit times and the transit time spread for single events
\item verification of the linear behaviour of the PMT signals up to high light intensities
\item determination of the photon detection efficiency (product of quantum efficiency and collection efficiency)
\item exploration of the afterpulse probability and their time characteristics
\item search for flashing PMTs
\end{list}

\begin{table*}[t]
     \centering
     \begin{tabular}{|c|c|c|c|c|}
     \hline
     Parameter & Min. & Typ. & Max. & Unit\\
     \hline
     Supply voltage for gain of 1 x $10^7$ & 1150 & 1500 & 1650 & V \\
     Anode dark count & - & 4000 & 8000 & 1/s \\
     Peak to valley ratio at SPE & 2.5 & 2.8 & - & - \\
     Transit time spread (FWHM) & - & 3.4 & 4.4 & ns \\
     After pulse probability (100 ns to 16 $\mu$s after main pulse) & - & - & 10 & $\%$ \\
     Quantum efficiency at 420nm & 22 & 25 & - & $\%$ \\
     \hline
     \end{tabular}
     \caption{PMT characteristics (and acceptance limits) at 25$^{\circ}$C and supply voltage for a gain of 1 x $10^7$ provided by Hamamatsu.}
     \label{PMTspecs}
\end{table*}

\noindent Besides the description of the test facility the following sections focus on the analysis and the results for the topics 1-6. The analysis of 7. and 8. is still ongoing and will be discussed in a separate publication.

\section{Test Facility}

The test facility for the above characterizations of the PMTs was built at MPIK in Heidelberg. The facility provides the possibility to measure 30 PMTs simultaneously during one calibration-run. During the whole calibration phase, three so called ``reference PMTs'' stayed in the setup and allowed to monitor the stability of the setup. Furthermore, the quantum efficiencies of these PMTs were absolutely calibrated by HPK which enables relative sensitivity measurements.
In this section we describe the setup, in particular the different light sources used, the electronics and the data acquisition.

\subsection{Setup of the Test Facility}
\label{SetupTestFacility}
The test facility was built inside a large 30 $\rm m^3$ steel box. This, in the following called ``Faraday room'', is well protected against external electromagnetic fields and is closed light-tight. To prevent undesirable light reflections, the inner walls have been covered with black fire-proof rag.

\begin{figure}[h]
\centering
\includegraphics[angle=270,width=\figwid]{\plots{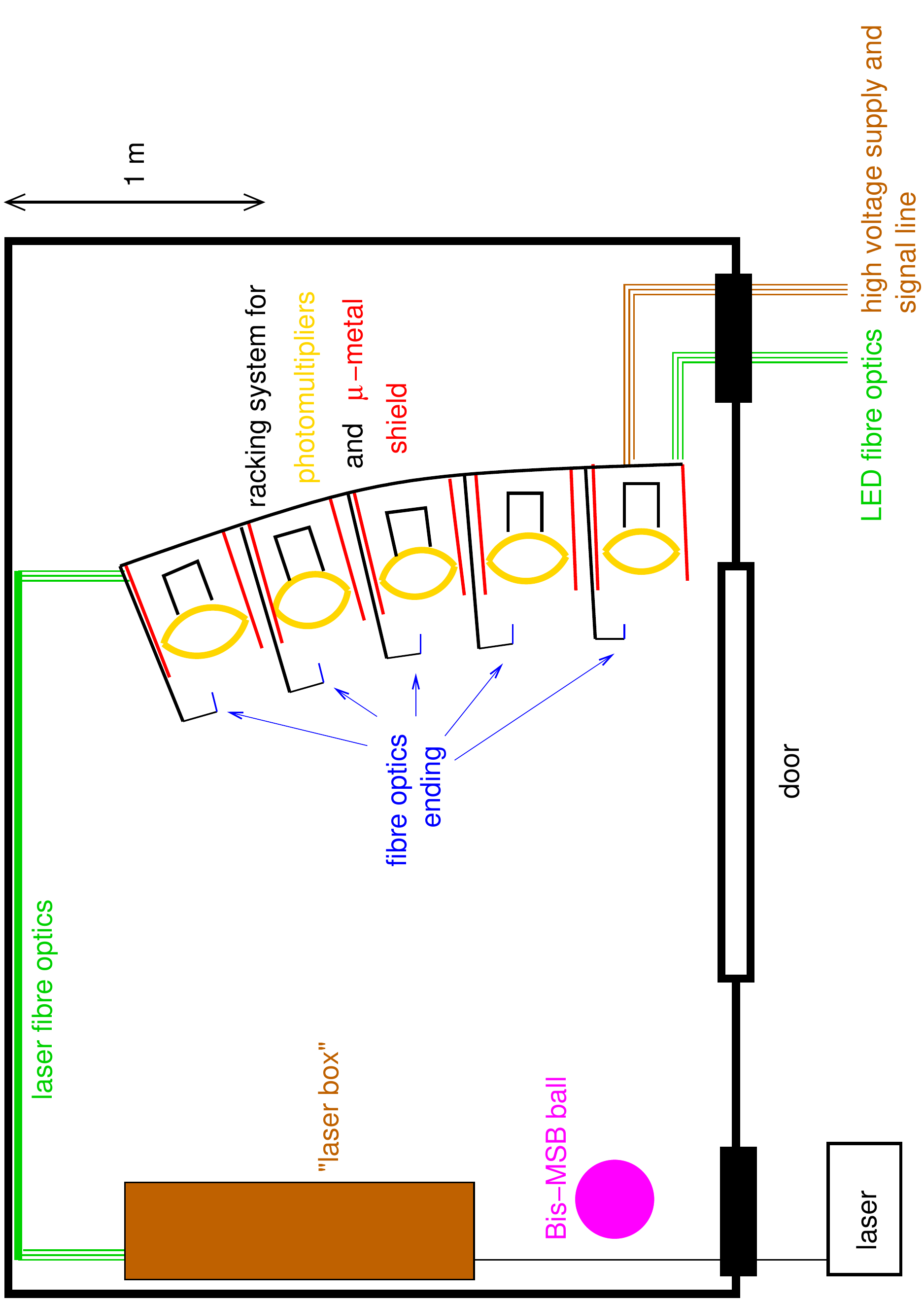}}
\caption{Topview of the Faraday dark room containing the racking system for 30 PMTs.}
\label{faraday}
\end{figure}

\begin{figure}[h]
  \centering
  \includegraphics[width=\figwid]{\plots{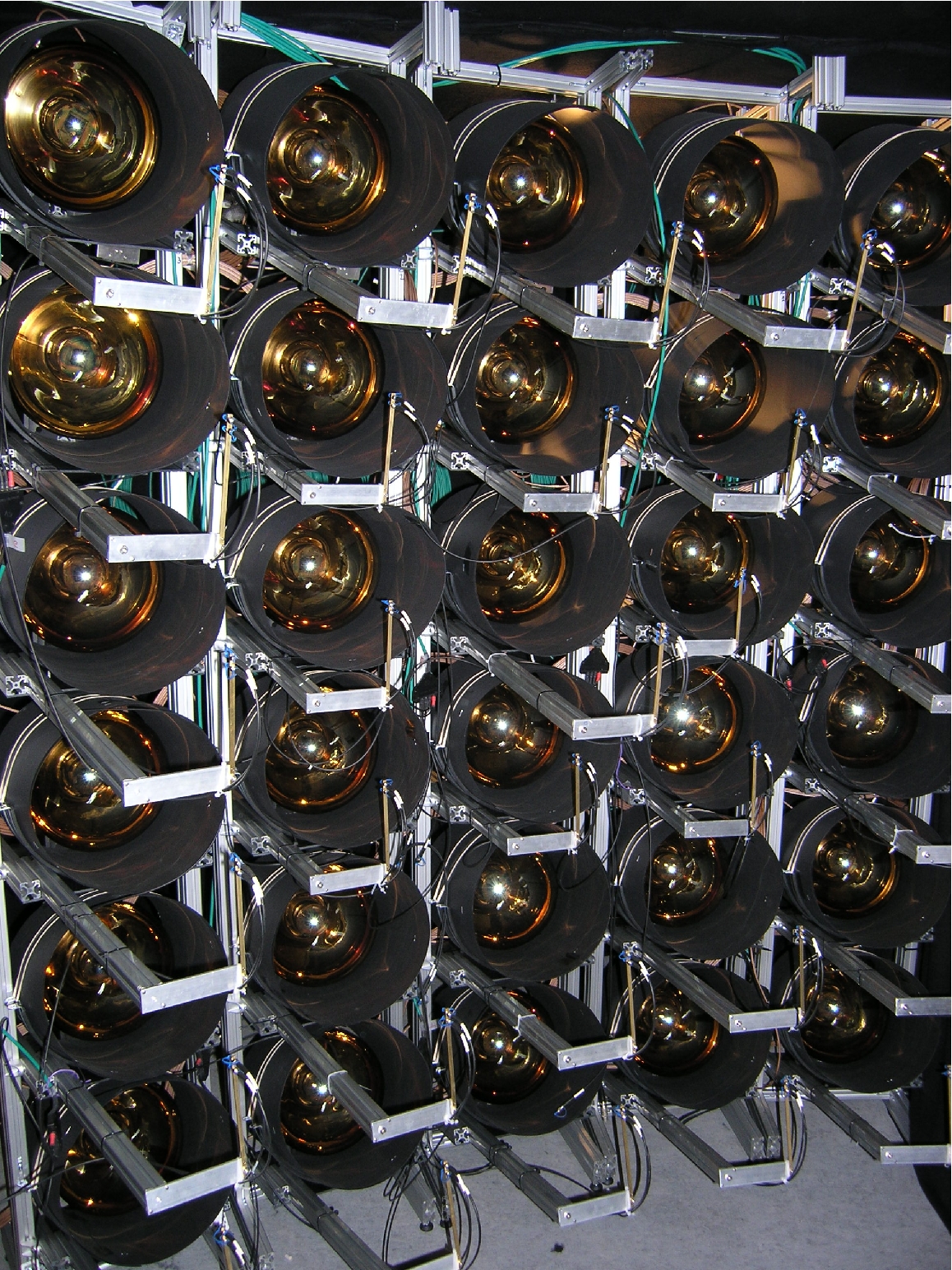}}
  \caption{The PMT racking system with mounted PMTs.}
  \label{PMTarray}
\end{figure}

Figure \ref{faraday} shows the topview of the Faraday 
room. The PMTs are mounted in slots of a racking 
system which consists of six rows and five columns (see Figure \ref{PMTarray}). As 
denoted in Figure \ref{faraday}, each slot is 
oriented towards the bis-MSB-ball, a central light source which allows one to illuminate all PMTs simultaneously. A metal arm at each slot feeds optical fibres coming from external light sources (see section \ref{light}) to illuminate each PMT individually.
The distance between the photocathode of a PMT mounted on the rack and the fibre attachment was chosen by two criterias  
with respect to the solid angle of the light spread at the end of the fibre. On one hand the photocathode should be fully illuminated, on 
the other hand light emitted from the fibre should not illuminate adjacent PMTs. Tests showed that both requirements were realized by a distance of about 50 cm between the end of the optical fibre and the photocathode.

For further magnetic shielding, a $\mu$-metal cylinder covered by black paperboard is mounted in each  slot. These cylinders have been produced for the use in the Double Chooz Experiment \cite{mumetal}.

\subsection{Light sources}
\label{light}

Three different light sources were used.
\begin{list}{$\bullet$}{
  \setlength{\topsep}{1ex}
  \setlength{\itemsep}{-0.3ex}
  \setlength{\leftmargin}{3.5ex}
}
\item \textbf{LED-boards:} These VME controllable 12-channel boards were developed by the electronic workshop at MPIK. The LEDs produce light at a wavelength of $\rm 380 \, nm$ with a light intensity that can be programmed for each channel individually from the SPE range up to a few hundred PEs. The light is transmitted by optical fibres to each PMT.
\item \textbf{Laser:} For measurements which require a high precision in timing we used a Picosecond Injection Laser (Advanced Laser Diode Systems A.L.S. GmbH) at a wavelength of $\rm 438 \, nm$ with a time jitter of $\rm 35 \, ps$. Inside the light tight laser box (see Figure \ref{faraday}) the light pulses are widened with a diffusor lens and shines on a perforated plate where optical fibres are connected which lead to the PMT rack. Because of this configuration, small differences of the light intensity are possible for different rack slots.
\item \textbf{Bis-MSB-ball:} This is a central light source of the test system. The glassblower workshop at MPIK built a quartz ball with a diameter of $\rm 11 \, cm$  which was filled with a bis-MSB/ethanol solution. Bis-MSB is a wavelength shifter which is also used for the scintillator of the Double Chooz experiment \cite{DCProposal}. The maximal spectral emission is at a wavelength of $\rm 420 \, nm$ and matches the optimal sensitivity of the PMTs \cite{PMT}. Four optical fibres are attached to guide LED signals via a quartz cylinder into the centre of the ball where the bis-MSB generates an isotropic light emission to illuminate the full PMT array. 
\end{list}

\begin{figure}[t]
  \centering
  \includegraphics[width=0.28\textwidth]{\plots{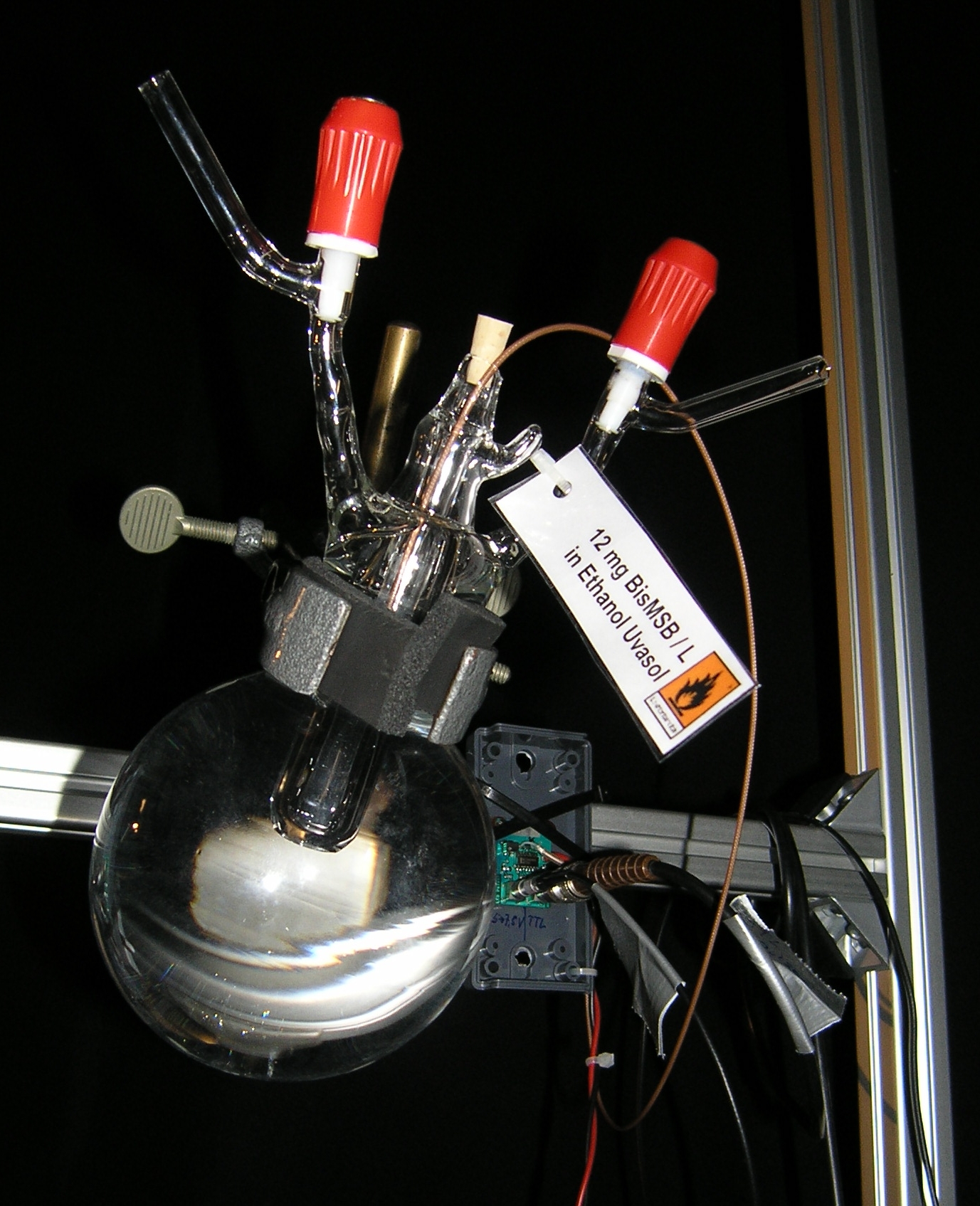}}
  \caption{The bis-MSB ball with attached optical fibres.}
  \label{bisMSB}
\end{figure}%

\subsection{Electronics and Data Acquisition System (DAQ)}
\label{DAQtext}

\begin{figure*}[t]
\centering
\includegraphics[angle=270,width=1.0\textwidth]{\plots{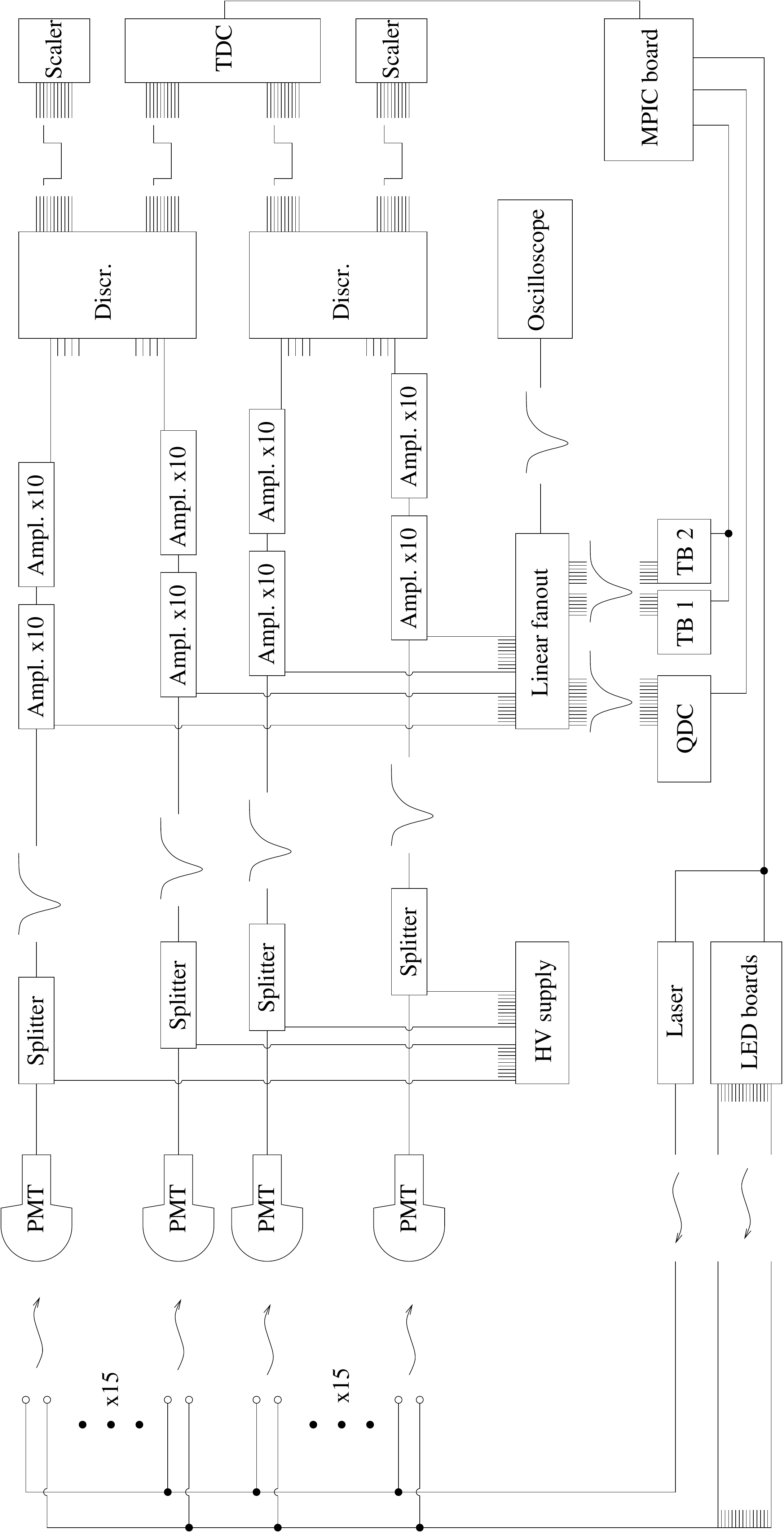}}
\caption{Schematic diagram of the electronics setup and the DAQ.}
\label{DAQ-sceme}
\end{figure*}%

The front end electronics and DAQ system is located outside of the Faraday room. The PMT cables are fed through the Faraday room's wall and are connected to splitter boxes where the signals are decoupled from 
the supply voltage. The splitter boxes are designed and produced by the CIEMAT group (Madrid) for the Double Chooz experiment. 
The high voltage supply is a SY2527 Universal Multichannel Power Supply System (CAEN) \cite{CAEN} which provides a $\rm 500 \, mV$ resolution for voltage setting and monitoring. 
The modules for processing and acquiring the PMT signals are based on standard NIM modules and a VME system, respectively. Figure \ref{DAQ-sceme} shows a schematic illustration of the data acquisition system.
  
In order to exploit the full dynamic range of the used acquisition modules and to provide the required logical signals, several NIM-modules are used:
\begin{list}{$\bullet$}{
  \setlength{\topsep}{1ex}
  \setlength{\itemsep}{-0.3ex}
  \setlength{\leftmargin}{3.5ex}
}
\item $4\times$ sixteen channel amplifier (fixed gain $\times 10$, Phillips Scientific (PS) \cite{PS} Mod. 776).
\item $4\times$ octal discriminator (threshold $\ge$ 10 mV, PS Mod. 710).
\item $4\times$ octal linear Fan-Out (PS Mod. 748).
\end{list}
The used acquisition modules are listed below:
\begin{list}{$\bullet$}{
  \setlength{\topsep}{1ex}
  \setlength{\itemsep}{-0.3ex}
  \setlength{\leftmargin}{3.5ex}
}
\item 32 channel charge-to-digital-converter (QDC V792, 12 bit resolution ($400 \, \rm pC$), CAEN \cite{CAEN})
\item 32 channel time-to-digital-converter (TDC V775, 12 bit resolution ($400 - 1200 \; \rm ns$),  CAEN)
\item $2 \times 16$ channel Scaler (V260E, 24 bit channel depth / $\rm 100 \, MHz$ counting rate, CAEN)
\item{The Double Chooz Trigger Board (DC-TB), RWTH Aachen \cite{BerndPhDT}}. 
\end{list}
All trigger signals, gate widths etc. for the measurements were provided by the MPIC-board, which was developed at the electronic workshop at MPIK. This board provides up to 15 logic signals with free programmable gate widths and delays.

\section{Measurements and results}
\subsection{Dark count rate behaviour}

Dark count events are caused by thermal emission of electrons from the photocathode which are not distinguishable from events caused by regular photoelectrons. The major part of these events are caused by single electrons. We measured the dark count rate with a discriminator threshold at 1/4 SPE pulse amplitude height (corresponding to  1 mV for an unamplified signal at a gain of $10^7$).

\begin{figure*}[t]
 \centering
  \includegraphics[angle=0,width=\doublefigwid]{\plots{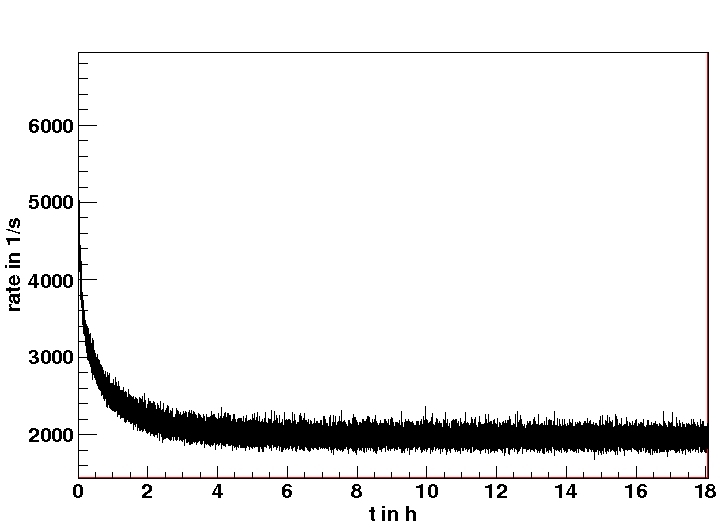}}
  \includegraphics[angle=0,width=\doublefigwid]{\plots{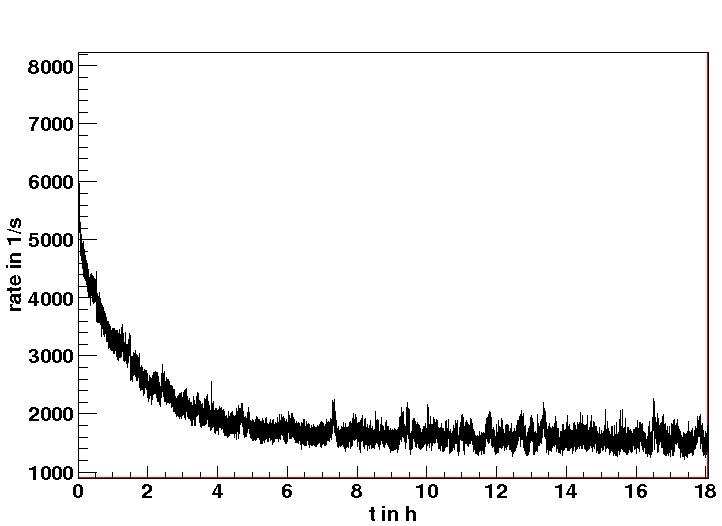}}
  \caption{Left: typical dark count rate behaviour during the first hours after installation. Right: example for an unstable dark count rate.}
  \label{DcR_example} 
\end{figure*}

If PMTs are exposed to light, which is unavoidable during the installation of the PMTs at the test setup, the dark count rate is increased during the first hours after turning on the high voltage. This phase of stabilization was monitored for a period of typically 20 hours with a scaler and the DC-Trigger-Board (see Figure \ref{DcR_example}, left). After this period of time, the rate has decreased to a reasonably stable value and all calibrations which will be described in the following sections were performed. Afterwards, the dark count rate was observed again for about one day to determine the final value and to inspect its stability.

The dark count rate depends on the temperature. During the measurements the temperature inside the laboratory at MPIK was stabilized by air conditioning at ($20 \pm 1)^\circ$C. We observed a change of the rate of roughly 100 c/s (counts/second)  per 1$^\circ$C temperature difference. Inside the Chooz underground laboratory the temperature is about 14$^\circ$C, therefore the dark count rate is expected to be lower in the final detector.

\begin{figure}[t]
\centering
\includegraphics[angle=0,width=\figwid]{\plots{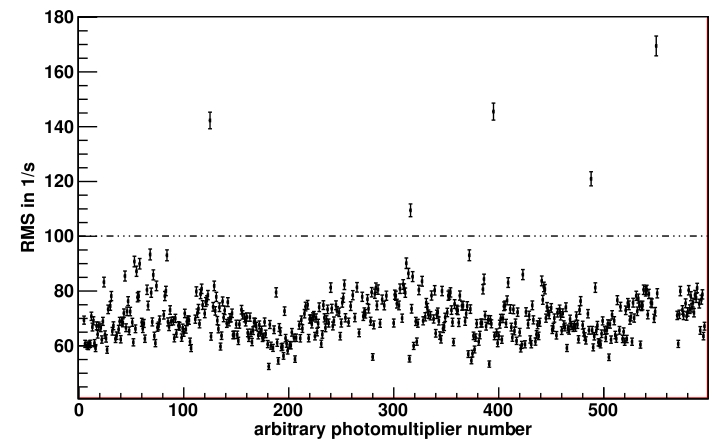}}
  \caption{RMS value from the last 20 minutes in the dark count rate measurement.}
  \label{DN_RMS} 
\end{figure}

The specification for the Double Chooz experiment is a dark count rate below $8000 \rm \; c/s$ per PMT.    
With one exception, all tested PMTs showed much smaller values with an average of $2200 \rm \; c/s$ at 20$^\circ$C with a standard deviation of $500 \rm \; c/s$.

To quantify the stability of the dark count rate, we calculated the root mean square (RMS) of taken data points during the stable phase (see Figure \ref{DcR_example}, right) to compare it with the value which is expected for statistical fluctuations. If the RMS was significantly higher (see Figure \ref{DN_RMS}) the PMT was not used for the Double Chooz detector.

\subsection{Gain Calibration}
\label{gaincali}

\begin{figure*}[t]
\centering
 \includegraphics[angle=0,width=\doublefigwid]{\plots{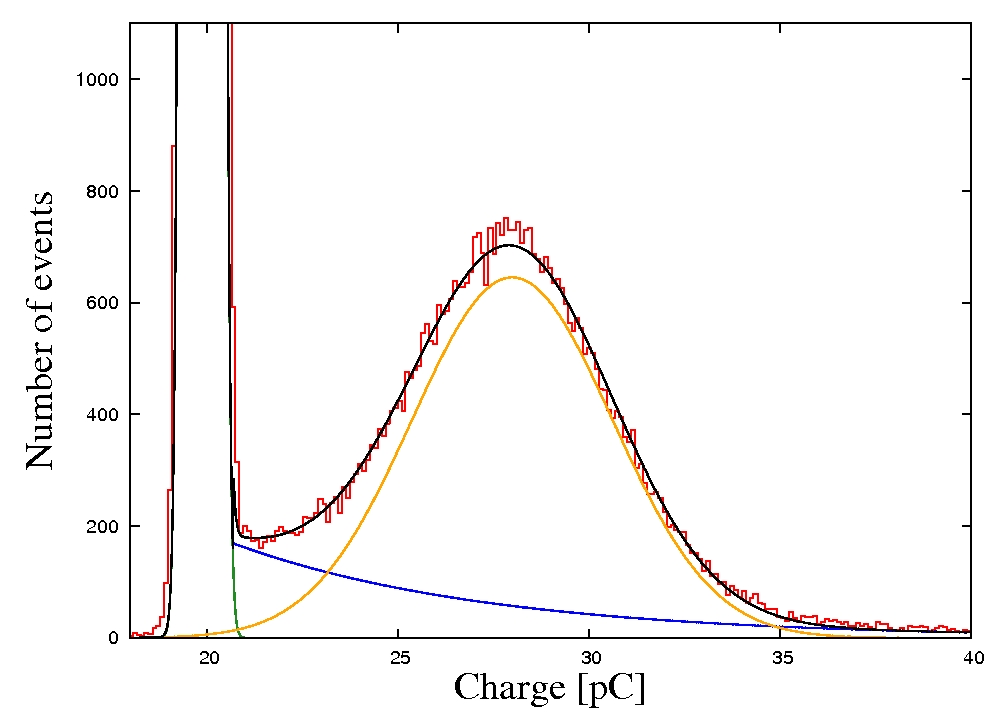}}
 \includegraphics[angle=0,width=\doublefigwid]{\plots{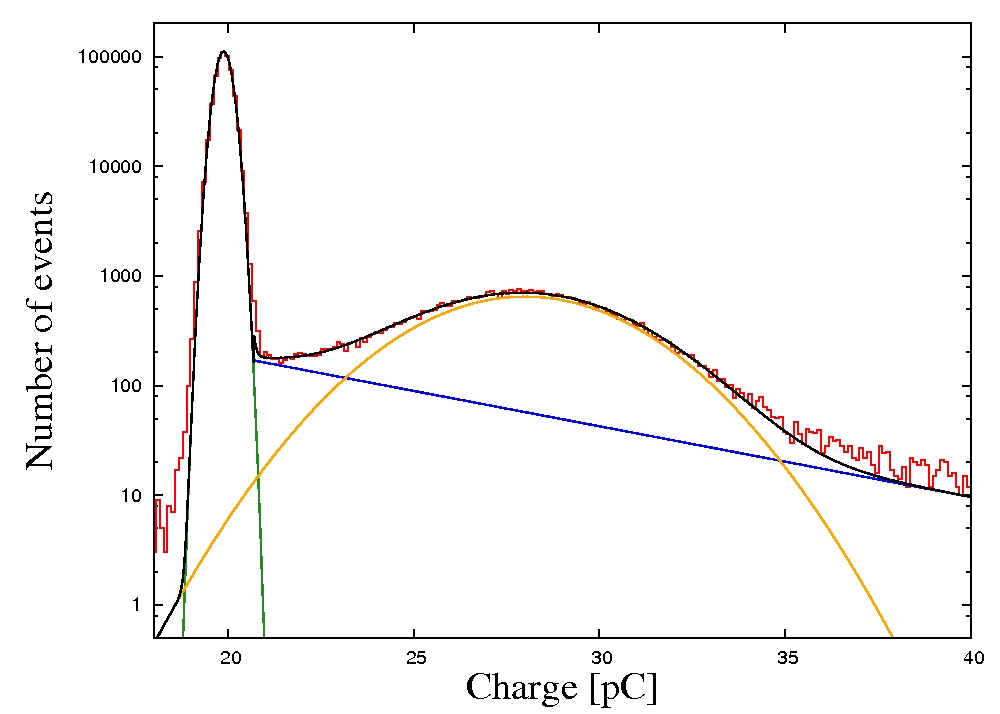}}
\caption{SPE charge spectrum in linear and logarithmic view, together with fit functions for pedestal and SPE peak. The red line shows raw QDC data. The pedestal peak was fitted by a Gaussian (green). The functions whose combination describe the SPE peak (see text) are shown in blue and orange. The black line is the combination of all fit functions which describes the complete spectrum.}
\label{QDCspec}
\end{figure*}

For this measurement, we used the LEDs at SPE level. The light intesities were adjusted in such a way that on average only one out of ten triggers gives a PMT signal. Assuming that the number of created photoelectrons per event $n_{\rm PE}$ is poisson distributed, the probability of $n_{\rm PE} \ge 2$ is only about 0.5\%. Therefore almost pure SPE spectra were observed. The total trigger rate of LED pulses was $\rm 1 \, kHz$. For each PMT, we took nine charge spectra (over $\rm 400 \, s$ each which corresponds to 400k events) at high voltage values ${\rm HV}_j \; (j=1, \, \ldots , \, 9)$ ranging between $\pm$ $\rm 200 \, V$ around the nominal high voltage $\rm HV_{\rm nom}$ (which is the value for a gain of $10^7$ as provided by HPK). The signal current arriving at the QDC was integrated within a time window of 200 ns. This leads to charge spectra for which an example is shown in Figure \ref{QDCspec}.
To analyse these spectra the pedestal peak was fitted by a Gaussian ${\rm Ped}(i)$ 
\begin{equation}
{\rm Ped}(i)= \frac{N_{\rm Ped}}{\sqrt{2\pi}\sigma_{\rm Ped}} \, \exp\left( -\frac{(i-\mu_{\rm Ped})^2}{2\sigma^2_{\rm Ped}}\right) \, .
\label{pedfitfkt}
\end{equation}
To describe the SPE peak a combination of an exponential and a Gaussian ${\rm SPE}(i)$ was used
\[
{\rm SPE}(i)=\frac{N_{\rm SPE}}{\sqrt{2\pi}\sigma_{\rm SPE}} \exp\left(-\frac{(i-\mu_{\rm SPE})^2}{2\sigma^2_{\rm SPE}}\right) + \] 
\begin{equation} 
\; + \frac{N_{\rm Exp}}{ \tau } \exp\left(-\frac{(i-i_{\rm min})}{\tau}\right) \, .
\label{spefitfkt}
\end{equation}
$N_{\rm Ped}$ and $N_{\rm SPE}$ are the number of entries in the pedestal and the SPE peak, respectively. $\mu_{\rm Ped}$ and $\mu_{\rm SPE}$ are the mean values of the fitted Gaussians, $\sigma_{\rm SPE}$ and $\sigma_{\rm Ped}$ their standard deviations, $N_{\rm Exp}$ and $\tau$ are the free fit parameters of the exponential which is an empirical function to describe bad amplified events, $i$ is the number of the QDC-bin and corresponds to a charge of $i \cdot 400/4096 \rm \, pC$, $i_{\rm min}$ is the endpoint of the pedestal fit and the starting point for the exponential fit which is roughly determined by the valley position.

\begin{figure}[h!]
  \centering
\includegraphics[angle=0,width=\figwid]{\plots{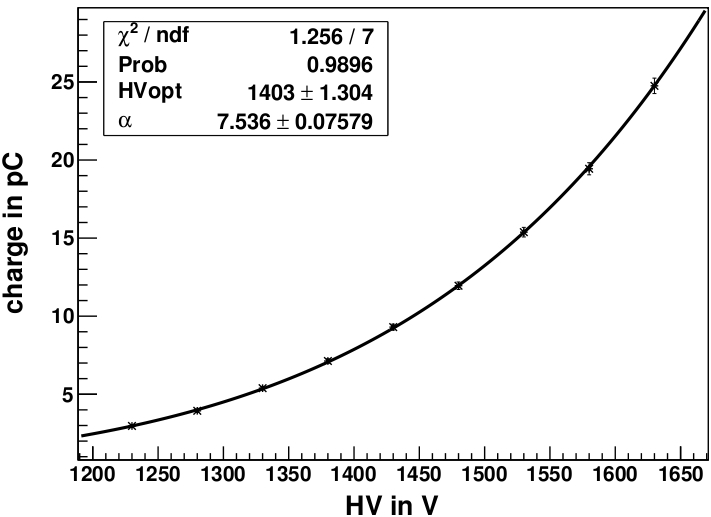}}
\caption{Example of a power law fit for determination of $\rm HV_{opt}$. The black points represent the charge $Q$ which was determined at a certain high voltage value according to equation (\protect \ref{AverageCharge}). The solid line is the fitted power law function equation (\protect \ref{HVfit}).}
\label{QvsHV}
\end{figure}

The fit functions and their combination are drawn in Figure \ref{QDCspec} showing model and data in good agreement.
From the results of both fits the average charge $Q$ corresponding to a SPE-signal is calculated
\begin{equation}
 Q = \mu_{\rm SPE}\,-\,\mu_{\rm Ped} \, .
\label{AverageCharge}
\end{equation}
For the nine different high voltage values ${\rm HV}_j$, the corresponding charges $Q({\rm HV}_j)$ are shown in Figure \ref{QvsHV}. The $Q({\rm HV})$ dependence is fit with a power law function 
\begin{equation}
 Q({\rm HV}) = 8.0 \, {\rm pC} \cdot \left(\frac{\rm HV}{\rm HV_{opt}}\right)^\alpha \; .
\label{HVfit}
\end{equation}%

The free fit parameters are $\rm HV_{opt}$ and $\alpha$, where $\rm HV_{opt}$ is the searched value for a gain of $10^7$ which corresponds to an average charge of $\rm 8.0 \, pC$\footnote{A gain of $10^7$ corresponds to a charge of $10^7 e = 10^7\cdot 1.602 \cdot 10^{-19}{\rm C} = 1.602 \, {\rm pC}$. An additional $50\Omega$ back termination of the PMTs causes a division by a factor of 2, finally an amplification ($\times 10$) in the electronics chain leads to a value of $8.0 \, \rm pC$.}.

\begin{figure*}[t]
\centering
\includegraphics[angle=0,width=7.8cm]{\plots{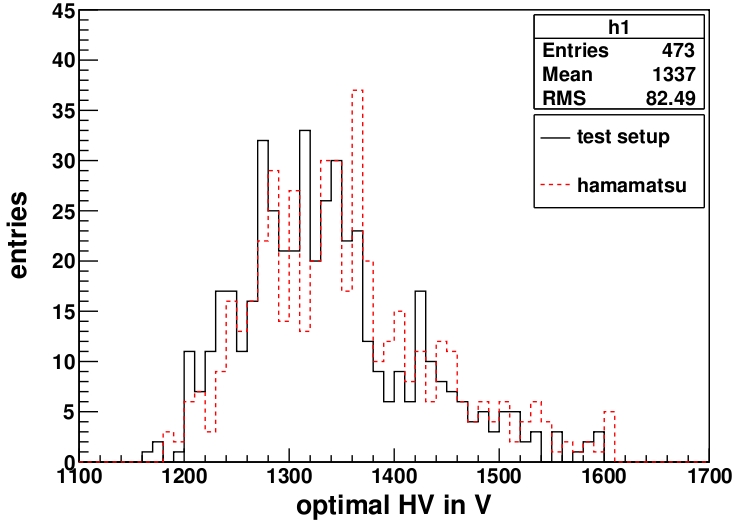}}
\hfill
\includegraphics[angle=0,width=7.8cm]{\plots{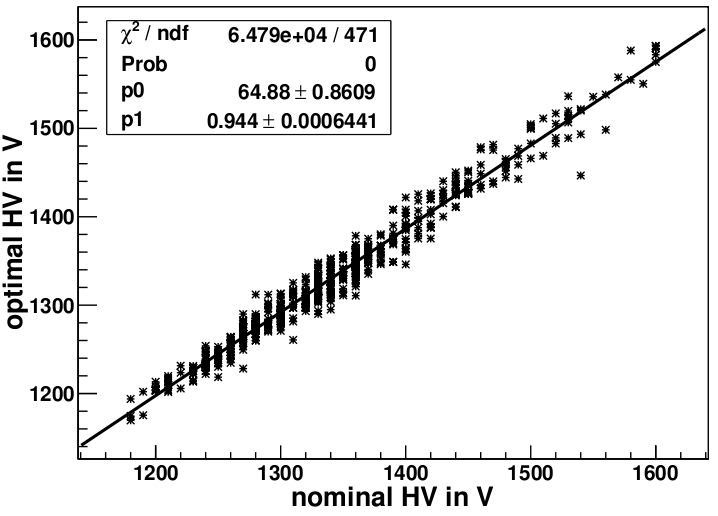}}
\caption{Left: Distribution of determined values $\rm HV_{\rm opt}$ (solid line) for a gain of $10^7$ compared to the values $\rm HV_{\rm nom}$ provided by HPK (dotted line) of all tested PMTs. Right: $\rm HV_{\rm opt}$ values versus the corresponding values $\rm HV_{\rm nom}$. Both design and actual HV values are confirmed to be consistent.}
\label{HVCorrelation}
\end{figure*}

\noindent The obtained results of ${\rm HV_{opt}}$ for all tested PMT are presented in a histogram shown in Figure \ref{HVCorrelation} (left), together with the values $\rm HV_{nom}$ provided by HPK. The good correlation between these values is shown in Figure \ref{HVCorrelation} (right).

\begin{figure}[t]
\includegraphics[angle=0,width=\figwid]{\plots{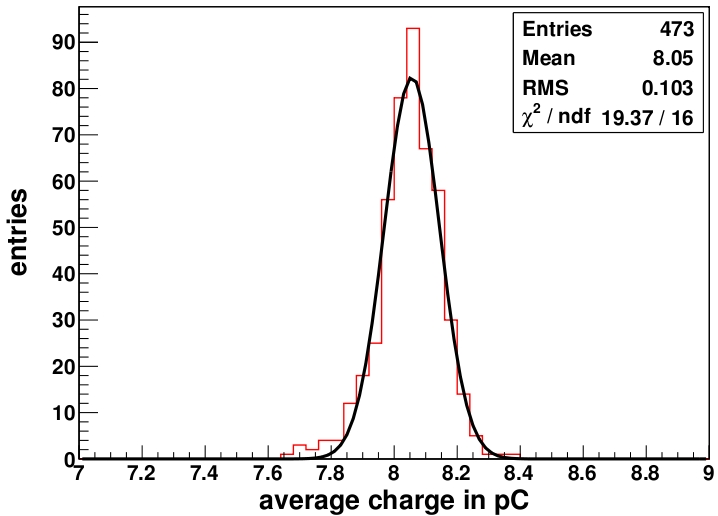}}
\caption{Average charge value of SPE signals at $\rm HV_{\rm opt}$.}
\label{Qhist}
\end{figure}

For the verification of the obtained results and a detailed characterization of the SPE response of each PMT an additional SPE charge spectrum was recorded at the previously determined $\rm HV_{\rm opt}$ value. The charge value for this spectrum is determined with the above procedure and is compared to the expected value of $\rm 8.0 \, pC$. The results for all PMTs are shown in Figure \ref{Qhist}.
The gain values deviate from the expected gain with a Gaussian distribution and relative RMS of $1.3\%$.
\begin{figure}[t]
  \includegraphics[angle=0,width=\figwid]{\plots{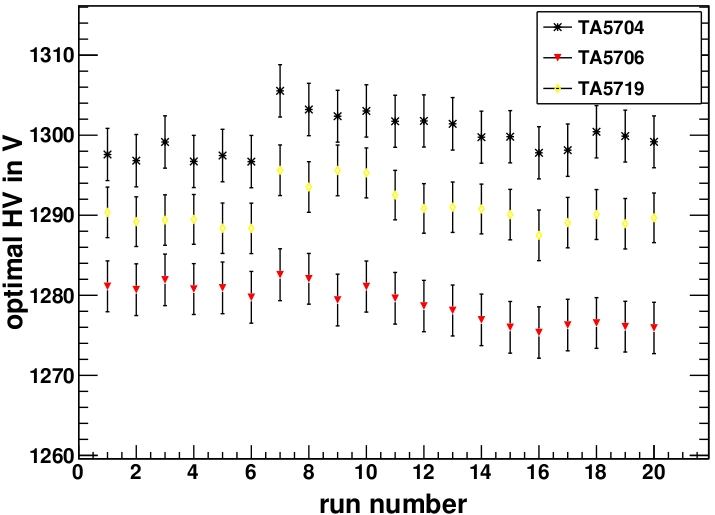}}
  \caption{${\rm HV}_{\rm opt}$ of the reference PMTs versus the run number.}
  \label{HVRefStab} 
\end{figure}
\begin{figure}[t]
  \includegraphics[angle=0,width=\figwid]{\plots{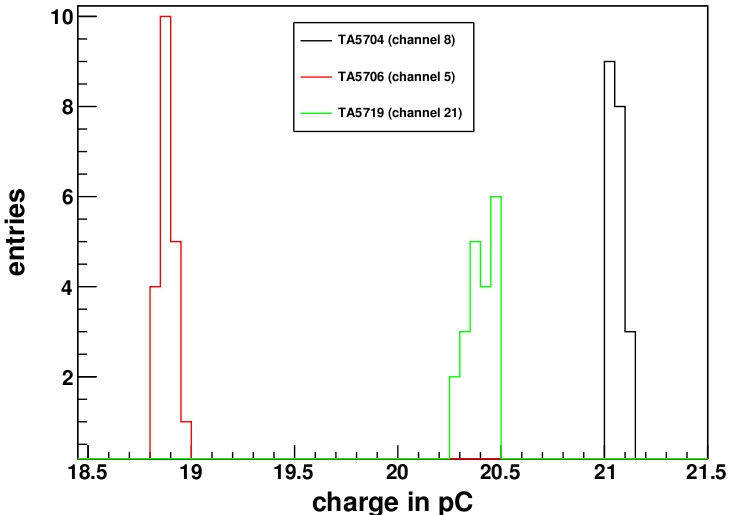}}
  \caption{Histogram of the determined pedestal values of the reference PMTs.}
  \label{HVPedStab} 
\end{figure}

Finally, we investigate the stability of the three reference PMTs which are repeatedly measured in each data run at fixed positions in the rack. Figure \ref{HVRefStab} shows the determined $\rm HV_{\rm opt}$ values as a function of the consecutive run numbers of successive PMT calibrations (corresponding to a period of about three months in total). The obtained values $\rm HV_{opt}$ agree within a few Volts, however a peculiar step is observable at run 7. Before this run, it was necessary to change some amplifier and fan-out channels which required re-calibrations for these channels. The visible step is consistent with an assumed systematic error of the calibrations of about 1\% (which was determined using a pulse generator).
In Figure \ref{HVPedStab} the determined pedestal values are shown in a histogram. The average relative spread of less than a half percent of the determined pedestal values shows a good stability of the test setup.


\subsection{SPE response - charge and transit time}

\begin{figure}[t]
\includegraphics[angle=0,width=\figwid]{\plots{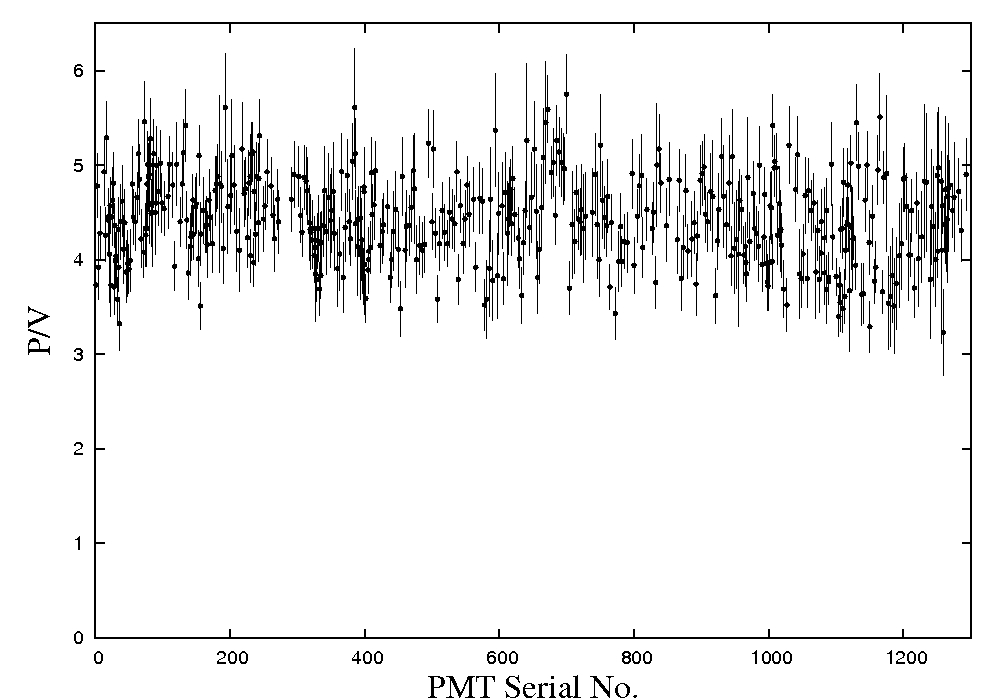}}
  \caption{Results for all tested PMTs regarding Peak-to-valley-ratio.}
  \label{P_V} 
\end{figure}

The PMT response for SPE pulses was characterized in detail using the high voltage $\rm HV_{opt}$. The DAQ provides the possibility to record simultaneously charge and transit times for each event. A typical SPE charge spectrum is shown in Figure \ref{QDCspec}. The charge resolution of the SPE peak is typically $\Delta E \approx 2 \, \rm pC$, or to give a relative number $\Delta E/E \approx 0.25$. The peak-to-valley-ratios $P/V$ (defined as the ratio between the maximum of the SPE peak and the minimum in the valley between the pedestal and the SPE peak) is in the range of $3.2 \le P/V \le 5.5$. The results  for all tested PMTs are shown in Figure \ref{P_V}. These numbers were improved substantially from previous smaller values after the introduction of the $\mu$-metal shields.

\begin{figure}[t]
\includegraphics[angle=0,width=\figwid]{\plots{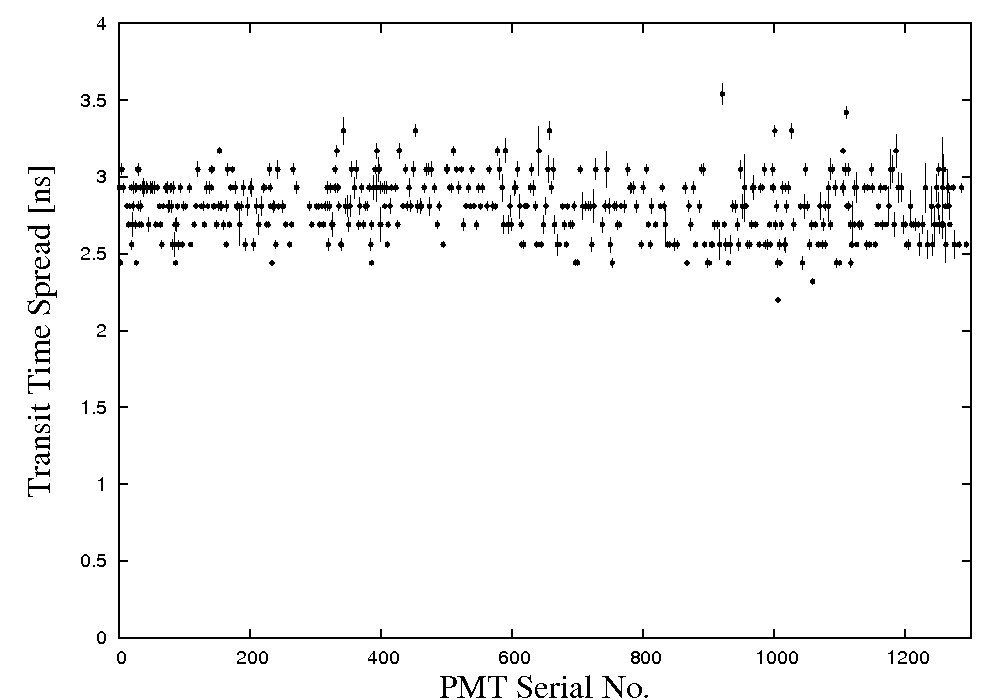}}
  \caption{Results for all tested PMTs regarding transit time spread.}
  \label{TTS} 
\end{figure}

\begin{figure*}[t]
  \centering
  \includegraphics[angle=0,width=0.48\textwidth]{\plots{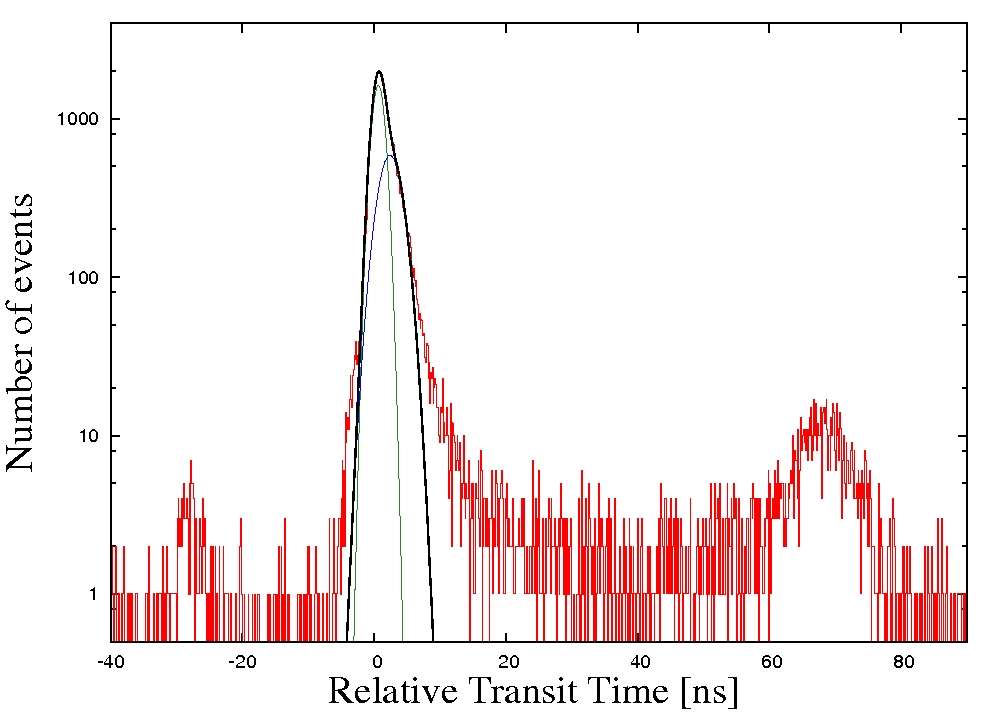}}
  \includegraphics[angle=0,width=0.48\textwidth]{\plots{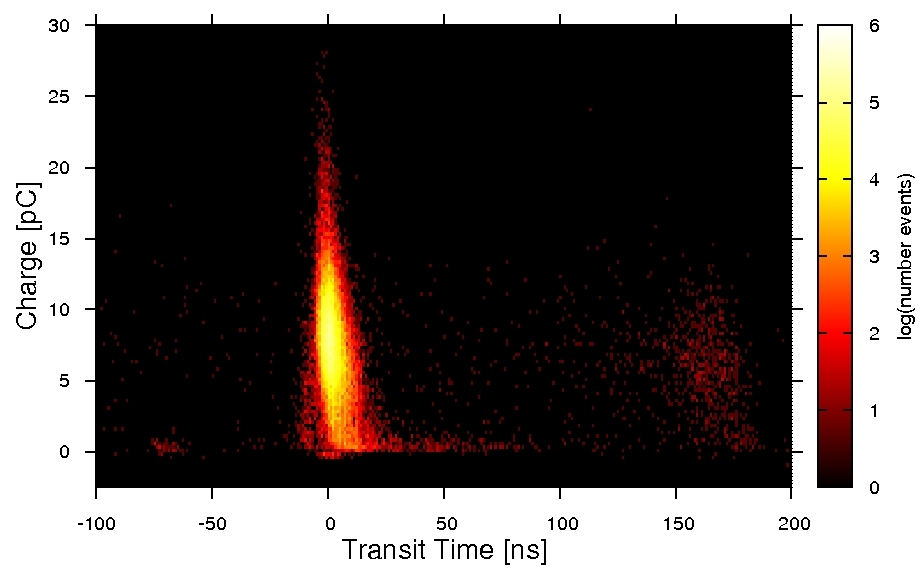}}
  \caption{Left: typical distribution of relative transit times including main-, pre- and late pulses. Right: charge versus transit time plotted event by event.}
  \label{TTS+scatter} 
\end{figure*}

The absolute transit times were measured including an additional offset related to the electronics and the used light source. The distribution of relative transit times is shown in Figure \ref{TTS+scatter} (left). Most of the events occur at roughly the same time after the time of the Laser trigger which defines $t=0$. This main peak can be described by the sum of two Gaussians. The transit time spread is defined as the width of this peak (full width at half maximum), the results of all tested PMTs are shown in Figure \ref{TTS}. The average transit time spread (FWHM) is $2.8 \, \rm ns$ with an RMS of $0.2 \, \rm ns$.

In addition to the main peak two more classes of events can be identified in the transit time distribution: pre- and late-pulses\footnote{Please note that the used single-hit TDC is not sensitve for afterpulses (caused by ionized residual gas) since only the first event after the light trigger is recorded.}. Pre-pulses are appearing about 30 ns before the main peak. They are presumably caused by photons which pass the photocathode and hit the first dynode producing a photoelectron. In this case one step of the electron multiplication is missed, and the charge of pre-pulses is lower than the SPE value. Late-pulses appear if photoelectrons are (in)elastically scattered at the dynode, reflected and re-accelerated. These late-pulses appear up to 80 ns later than the main pulses and have a slightly smaller charge than regularly accelerated photoelectrons. The charge-transit time correlation is shown in Figure \ref{TTS+scatter} (right). The probability of pre-and late-pulses are roughly 0.1\% and 3\%, respectively.

\subsection{Verification of linear behaviour up to high light intensities}
\label{linearbehave}

The expected light levels of neutrino events in the Double Chooz experiment are typically in the range of a few PE. Nevertheless, signals induced by cosmic muons, fast neutrons etc. may exceed the SPE region substantially. Therefore we studied the linear behaviour of the PMTs up to light intensities corresponding to 30 photoelectrons (PE) (in a few cases, up to 300 PE). 

\begin{figure}[h]%
\includegraphics[angle=0,width=\figwid]{\plots{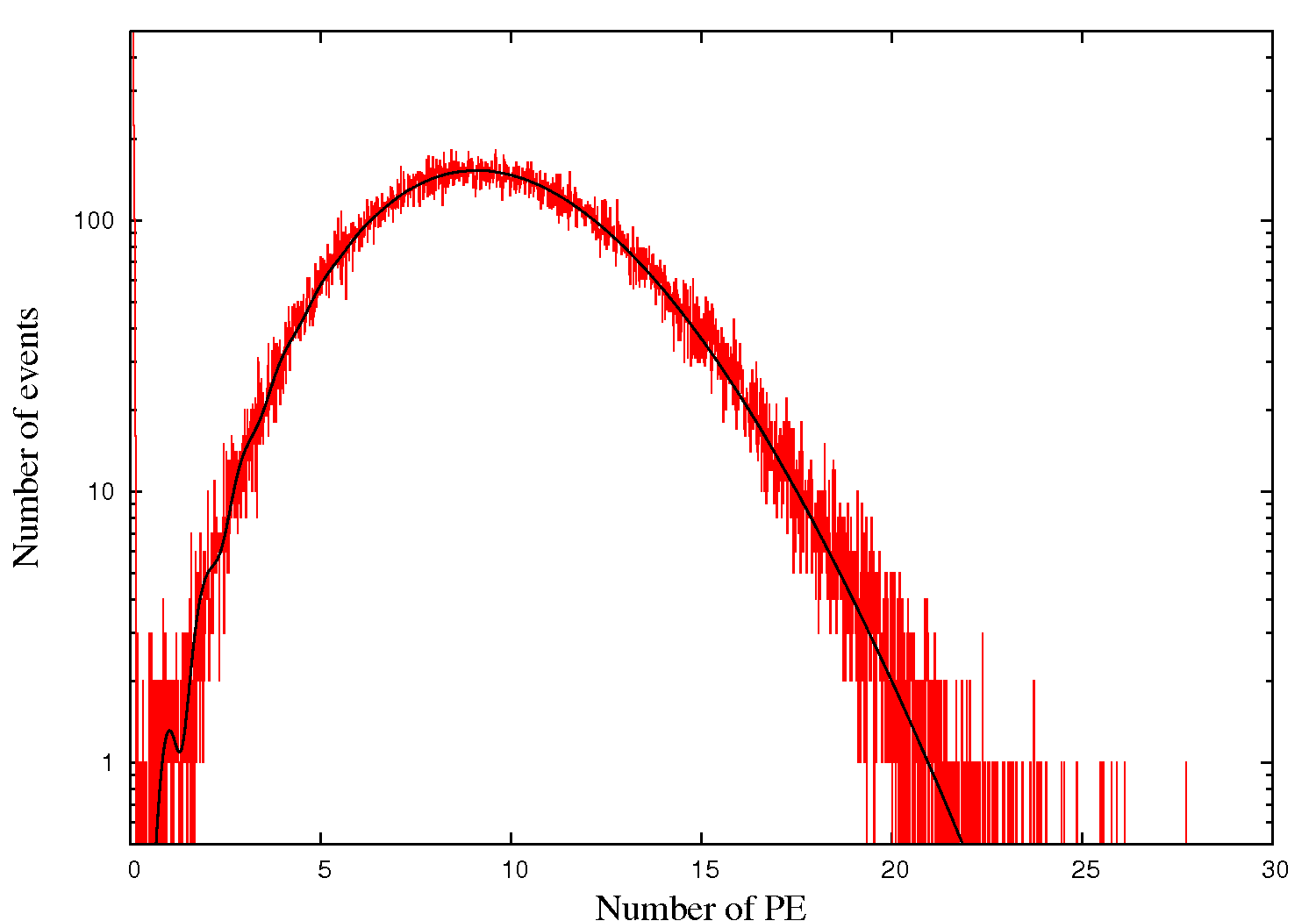}}
\caption{Typical charge spectrum for light intensities in the multi-PE region. The data (red) were fitted by the function described in equation (\protect \ref{poissonfitfkt}) (black).}
\label{QDCpoiss}
\end{figure}

The measurement proceeds similar to the gain calibration except for the use of the bis-MSB-ball as central light source. Because of the higher light intensity almost every pulse of the bis-MSB-ball leads to PMT signals. Therefore, additional DAQ triggers without light pulses were taken for a precise pedestal determination.
A typical charge spectrum is shown in Figure \ref{QDCpoiss}. The pedestal was fitted with a Gaussian (see equation (\ref{pedfitfkt})), and the spectrum was corrected for this pedestal. The charge is converted into the corresponding number of PEs, using the values determined during the gain calibration (equation (\ref{spefitfkt})). The expected discrete Poissonian distribution with the mean value $\mu$ is smeard by the SPE-resolution $\sigma$. Therefore a convolution of the Poissonian with a Gaussian was used to fit the multi-PE data (MPE) and to 
determine $\mu$ 
\[ {\rm MPE}(i)= N \sum_{k=1} \left( \frac{\mu^{k}}{k!} \exp(-\mu) \times \right. \qquad \quad \quad \]
\begin{equation}
\qquad \qquad \qquad \left. \times \frac{1}{\sqrt{2 \pi k} \sigma} \exp \left(- \frac{(i-k)^2}{ 2 k \sigma^2}\right) \right) \, .
\label{poissonfitfkt}
\end{equation}
The free fit parameter is $\mu$ which is the average number of photoelectrons. $N$ is the total number of entries and $\sigma$ the SPE resolution. $i$ is the number of the bin of the QDC corresponding to a charge calibrated in PE.

\begin{figure}[t]
\centering
\includegraphics[angle=0,width=\figwid]{\plots{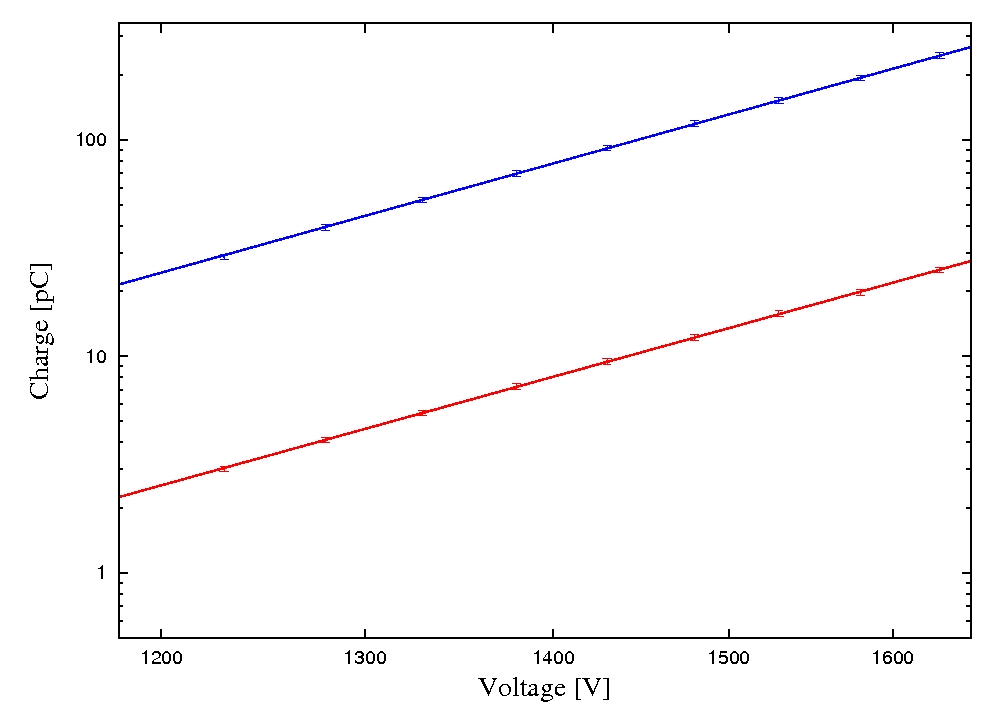}}
\caption{Average charge as a function of corresponding HV values at two different light intensities in double-logarithmic scale.}
\label{LinScan}
\end{figure}
\begin{figure}[t]
\centering
\includegraphics[angle=0,width=\figwid]{\plots{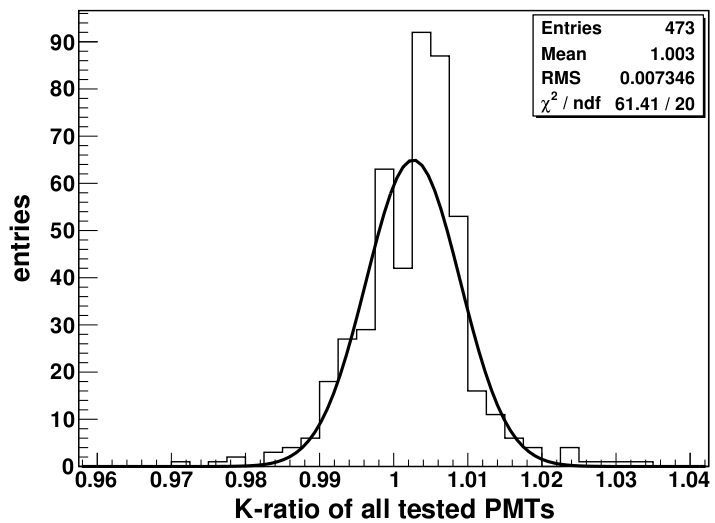}}
\caption{Distribution of the $K$ ratios of all tested PMTs and fitted Gaussian.}
\label{Khist}
\end{figure}

For each PMT, nine charge spectra for the same ${\rm HV}_j$ values as for the gain calibration in section \ref{gaincali} are measured. The determined $\mu_j$ are converted to the corresponding charges $Q_j$ which are plotted as function of the ${\rm HV}_j$ values in Figure \ref{LinScan}. A power law function is fitted to the data and compared to the gain calibration (equation (\ref{HVfit}))
\begin{equation}
 Q({\rm HV}) = I \cdot 8.0 \,\rm{pC} \cdot \left(\frac{\rm HV}{\rm HV_{\rm opt}}\right)^{\alpha} \, .
\label{linfit}
\end{equation}
The free fit parameters are the relative number of PE for this light level $I$ and the exponent $\alpha$. $\rm HV_{\rm opt}$ had been determined earlier during the gain calibration (see section \ref{gaincali}).
In Figure \ref{LinScan} both HV scans are shown in double logarithmic scale. The slope $\alpha_{\rm MPE}$ for higher light intensities is similar to the slope $\alpha_{\rm SPE}$ at small light levels indicating a good linear response. A good linear behaviour should lead to a ratio 
$ K\,=\,\frac{\alpha_{\rm MPE} } {\alpha_{\rm SPE}}$ close to unity. 
The resulting values of $K$ for all tested PMTs are shown in Figure \ref{Khist}. The mean of the distribution is close to 1 with a 
spread of only 0.6$\%$. No indications of non-linear behaviour up to signals corresponding to 30 PE were observed for any PMT.

For a few PMTs, further studies of the linear response have been performed after the completion of the calibration measurements with even higher light intensities. In order to avoid saturation of the QDC one amplifier in the DAQ chain has been removed. The results lead to the conclusion of a good linear response up to 250-300 PE when operating the PMTs at $\rm HV_{\rm opt}$.

\subsection{Determination of the photon detection efficiency}
\label{Sensitivity}

As an important characteristic of the PMTs, we determined the photon detection efficiency $\varepsilon$ in order to identify bad PMTs. The efficiency $\varepsilon$ is defined as the product of the quantum- and the collection efficiency $\varepsilon= QE \times CE$. The quantum efficiency ($QE$) indicates the ratio of the number of created PEs and the number of photons hitting the photocathode. The collection efficiency ($CE$) determines the number of PEs which reach the first dynode. With our setup it was not possible to measure both quantities ($QE$, $CE$) independently, however from an experimental point of view only the product $\varepsilon$ is of interest. Following the informations given by the manufacturer, the collection efficiency is 0.9 and the quantum efficiency about 0.25 in the region of interest at a wavelength of 420 nm.  The photon detection efficiency $\varepsilon$ depends on the voltage between the photocathode and the first dynode, the structure of the electrical field as well as on the effective area of the first dynode.

The bis-MSB-ball is used to illuminate the complete PMT array and the absolute number of photons reaching the cathodes is unknown. Therefore, the performed calibrations are relative, comparing each PMT with the quantum efficiencies of three absolutely calibrated reference PMTs provided by HPK.

Because the light source is not perfectly stable in time, the light intensity of each event was monitored by a fixed reference PMT which remained in the same slot during all measurements. Furthermore, the geometric dependencies were calibrated by moving two additional reference PMTs subsequently slot by slot throughout the complete array. For each slot $s$ a geometry factor $g_s$ was determined, using the ratio of the mean values of the fixed PMT $\mu_{\rm fix}$ and the reference PMT at the corresponding slot $\mu_s$, defining $g_s := \mu_{\rm fix}/{\mu_s}$. For both reference PMTs (${\rm ref} = 1, 2$) the procedure was repeated for two different light intensities $I_{1,2}$. The final calibration for each slot was obtained by averaging
\begin{equation}
g_s^{\rm ref}=\frac{1}{2} \cdot {\left( \frac{\mu_{\rm fix}^{I_1}}{\mu_{s}^{I_1}} + \frac{\mu_{\rm fix}^{I_2}}{\mu_{s}^{I_2}} \right)} \; .
\label{Gplace}
\end{equation}
By comparing the response $\mu_{\rm PMT}$ of a PMT at slot $s$ with the simultaneous response $\mu_{\rm fix}$ of the fixed PMT it is possible to evaluate the value $\varepsilon_{\rm PMT}$ of the unknown PMT relative to the reference PMTs used during the geometric calibration
\begin{equation}
 \varepsilon_{\rm PMT} = \frac{\mu_{\rm PMT}}{\mu_{\rm fix}} \, g_s^{\rm ref} \,  \varepsilon_{\rm ref} \; .
\label{sensi}
\end{equation}	
To increase statistics, the measurements were done at three different light intensities in the multi-PE region. In addition (as mentioned above), we used two independent reference PMTs to reduce  systematic errors regarding the reference QE and in order to have a backup solution if one of the reference PMTs should show any unexpected irregularities (which did not happen). As final value we used the arithmetic mean of the six results (three intensities and two reference PMTs).
For all charge spectra the same analysis methods were used as described in section \ref{linearbehave}.

\begin{figure}[t]
\centering
\includegraphics[angle=0,width=\figwid]{\plots{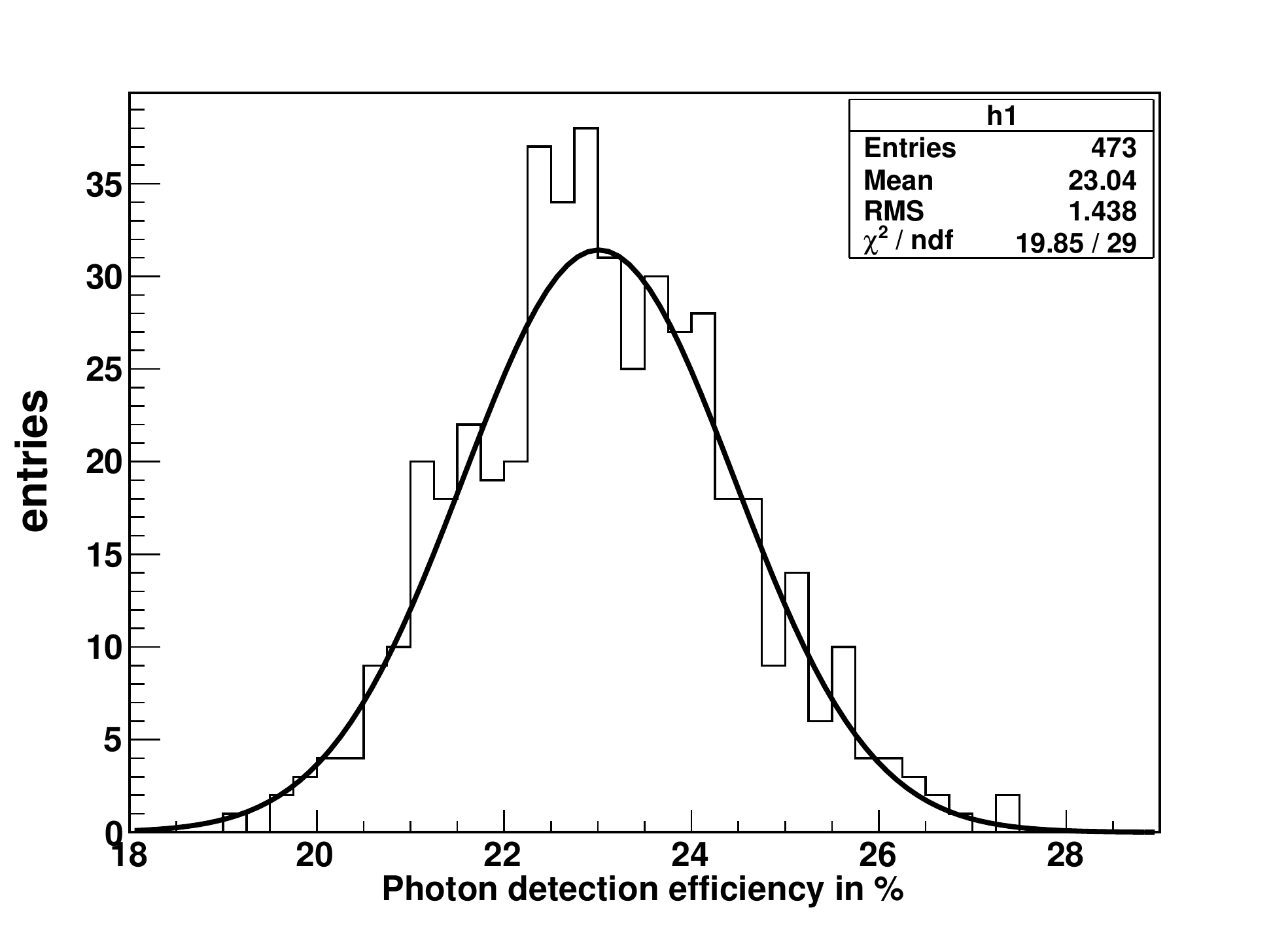}}
\caption{Photon detection efficiency $\varepsilon = CE \times QE$ of all tested PMTs and fitted Gaussian.}
\label{S}
\end{figure}

The geometrical calibration procedure which leads to equation (\ref{Gplace}) performed at the begin and was repeated twice during the calibration phase to control the stability of the system. It turned out that the used light source (the bis-MSB-ball) was not perfectly stable with respect to geometrical anisotropies. This means the values $g_s^{\rm ref}$ were affected by changes in time, either slowly or stepwise which could not be determined by a backdated analysis. These shifts were probably caused by small changes of the position of the optical fibers leading light from the LED boards inside the crystal ball. Analysing the response of the two reference PMTs with respect to the fixed PMT   (all these PMTs were present during the complete period of calibration) and comparing these values with the results of the three geometrical calibrations we estimated the relative systematical error of the photon detection efficiency  to $\Delta\varepsilon \le 5\%$. This precision is lower than expected before the calibration campaign, however it is sufficient to identify bad PMTs which was the main goal of these measurements. In total, there were only 3 PMTs with a photon detection efficiency lower than 19\% which were not dedicated for installation in the Double Chooz detector. 

\begin{figure}[t]
\centering
\includegraphics[angle=0,width=\figwid]{\plots{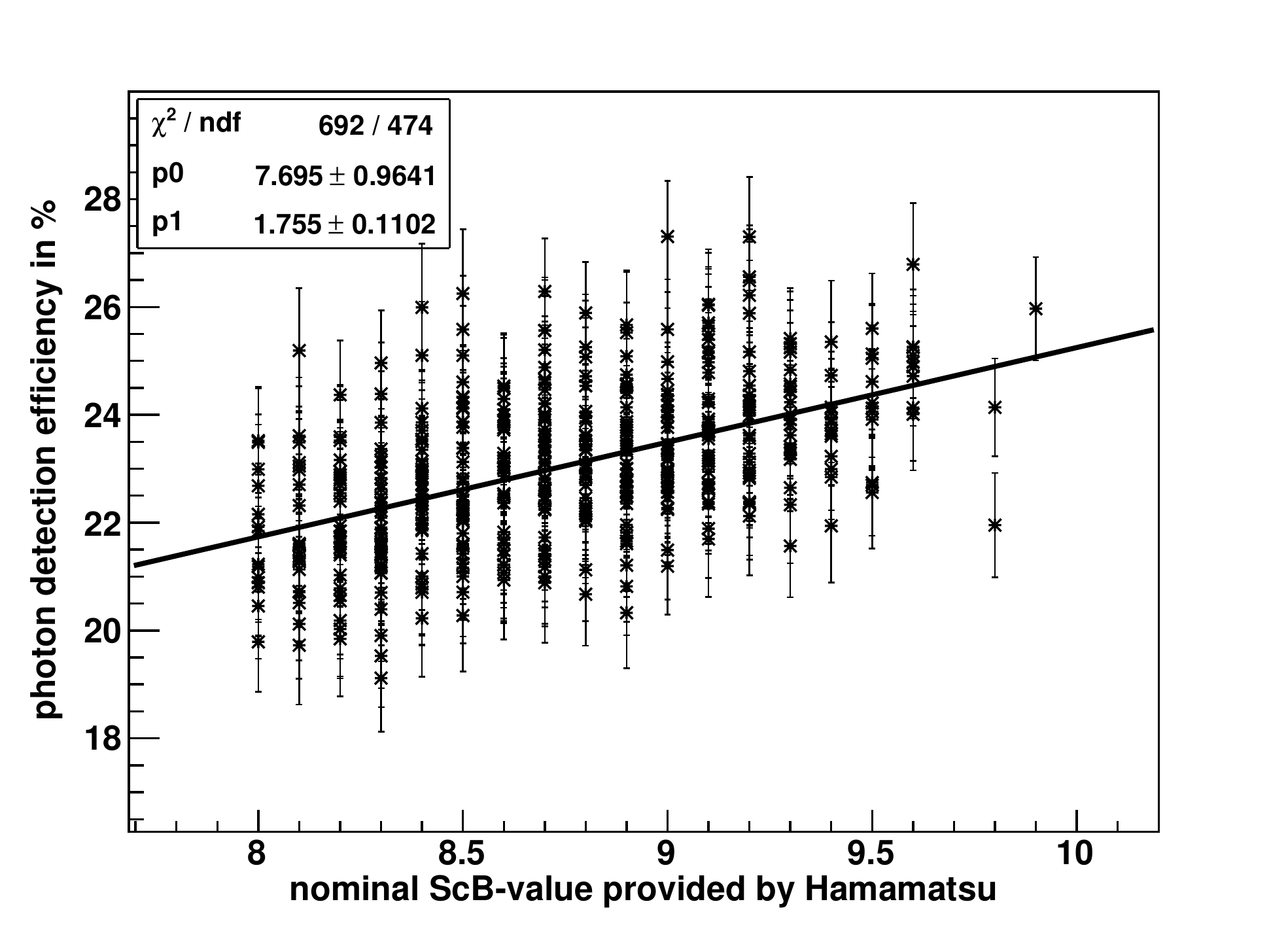}}
\caption{Photon detection efficiency $\varepsilon$ versus the corresponding ScB-value.}
\label{ScB}
\end{figure}

HPK provided a  value describing the sensitivity for blue light (ScB = sensitivity cathode blue index) for each PMT, which is expected to be correlated to the determined efficiency values $\varepsilon$. In Figure \ref{ScB} the results $\varepsilon$ of all PMTs are plotted versus their ScB-value. The correlation is obvious, although not very strong which is  caused by the uncertainties of both values and differences in the collection efficiencies which are included in $\varepsilon$.

\section{Conclusions}
For the Double Chooz experiment, 474 photomultiplier tubes were tested and calibrated regarding the dark count rate, gain as a function of high voltage, SPE response, transit time spread, linear behaviour and photon detection efficiency. One PMT showed no signal at all, another one lost its vacuum after the calibration measurements for an unknown reason. For 5 PMTs we noticed an unstable dark count rate. In addition, one of them showed an inacceptable dark rate of more than $10000\rm \; c/s$ . For 3 PMTs we determined an  photon detection efficiency lower than 0.2. In total 8 PMTs are not dedicated to be installed inside the Double detectors and will be kept as spare and to perform additional offsite tests in the future.     
    
\section*{Acknowledgments}
This work is supported by the DFG (Deutsche Forschungsgemeinschaft). 

We thank the whole Double Chooz PMT group for the excellent cooperation and would like to point out particularly the important contributions during the PMT calibration phase
of the CIEMAT group (Madrid, Spain) and the Double Chooz Japan group (Hiroshima Institute of Technology, Kobe University, Niigata
University, Tokyo Institute of Technology, Tokyo Metropolitan
University, Tohoku Gakuin University, Tohoku University).


\end{document}